\documentclass[showpacs,prl,preprintnumbers,amsmath,amssymb,aps,floatfix,twocolumn]{revtex4}
\usepackage{graphicx}
\usepackage{dcolumn}
\usepackage[sort&compress]{natbib}
\usepackage{bm}
\usepackage[dvips]{epsfig}

\begin{document}


\title{Thermal (in)stability of type I collagen fibrils.}
\author{S.G. Gevorkian$^{1}$, A.E. Allahverdyan$^{1}$, D.S. Gevorgyan$^{2}$, A.L. Simonian$^{3}$}

\address{$^{1}$ Yerevan Physics Institute, Alikhanian Brothers St. 2, Yerevan 375036, Armenia.\\
$^2$ Yerevan State Medical University, Koryun St. 2, Yerevan, 375025, Armenia.\\
$^{3}$ Materials Research and Education Center 275
Wilmore Auburn University, Auburn AL 36849-5341 USA.}

\begin{abstract}

We measured Young's modulus at temperatures ranging from $20$ to $100
^{\circ}$C for a collagen fibril taken from rat's tendon.  The hydration
change under heating and the damping decrement
were measured as well.  At physiological temperatures $25-45^{\circ}$C
Young's modulus decreases, which can be interpreted as instability
of collagen. For temperatures between $45-80^{\circ}$C
Young's modulus first stabilizes and then increases with decreasing the
temperature.  The hydrated water content and the damping decrement have
strong maxima in the interval $70-80^{\circ}$C indicating on complex
inter-molecular structural changes in the fibril.  All these effects
disappear after heat-denaturating the sample at $120^\circ$C.  Our main
result is a five-stage mechanism by which the instability of a single
collagen at physiological temperatures is compensated by the interaction
between collagen molecules within the fibril.

\end{abstract}


  

\maketitle

Type I collagen is the major structural element in the extra-cellular
matrix.  The native state of collagen is made up by three polypeptide
chains, which are twisted together into a triplex \cite{rama}.  Naively,
the collagen triplex is expected to be very stable, since it forms
fibrous connective tissues of bones, skin and tendons; see Fig.~1 for
the tendon hierarchical structure.  But in contrast, the collagen
triplex denaturates into separate chains at the helix-coil transition
$T_{\rm hc}$, which for mammals and birds is close to
\cite{Priv,Arno}|or even lower from \cite{Leikina}|the body temperature
(for poikilotherms $T_{\rm hc}$ relates to the upper environmental
temperature). Much attention was devoted to the physiological meaning of
this marginal thermal (in)stability \cite{Arno,Leikina,Persikov}, which
is generally explained as compromising between the instability of a
single collagen versus flexibility of collagen fibrils
\cite{Leikina,Persikov}.  However, it is so far not clear by which
specific mechanisms marginally instable collagen molecules achieve to
form the stable collagen fibril. 

We approach to this problem by studying thermal denaturation of a
collagen fibril via its mechanical characteristics such as Young's
modulus, logarithmic decrement of damping, and the hydrated water
content. The Young's modulus of collagen triplexes and fibrils was
measured via various methods \cite{buehler,gaut_buehler,young}. The
hydrated water was found to be essential for maintaining the triplex
\cite{water,water1}. It is also believed to be important for the
structure of fibril, though direct experimental evidences for this are
lacking \cite{water1}. Combining these quantities enables us to contrast
features of a single collagen molecule to those of a fibril, and study
the impact of the inter-molecular interaction on the fibril stability.
The structural transformations in the fibril are determined by
competition between entropy increase versus the intermolecular
interactions \cite{gaut_buehler}. It is known that for such materials
the mechanical properties can provide information not only on internal
elastic forces, but also on the molecular processes involving stress
relaxation \cite{Starikov}. In particular, these properties can uncover
relaxation processes both in the main polymer chain and side groups,
e.g., they revealed the phenomenon of low temperature glass transition
in the surface layers of protein molecules \cite{Mor1}. 

\begin{figure}
\includegraphics[width=250pt]{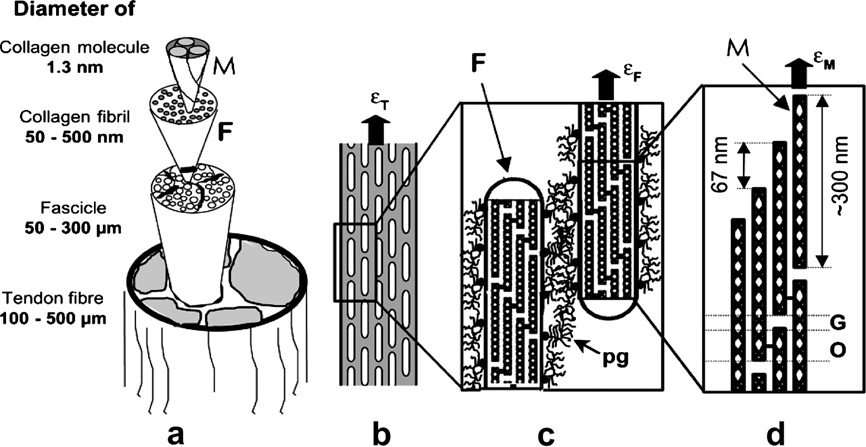}
  \caption{\label{fig1}
({\bf a}) Hierarchic tendon structure: collagen triplexes (M), fibrils (F),
fascicle and fiber.  ({\bf b}) Fibre is composed of fascicles. 
({\bf c}) The fascicle is a composite of collagen
fibrils in a proteoglycan-rich matrix (pg).  ({\bf d}) The fibril structure:
triplehelical collagen molecules (M) are staggered with an axial
spacing of $67$ nm. There is a
succession of gap (G) and overlap (O) zones. The triplexes are
stabilized in the fibril by intermolecular cross-links, direct hydrogen bonds and
water-mediated hydrogen bonds.  $\varepsilon_{\rm T}$, $\varepsilon_{\rm
F}$ and $\varepsilon_{\rm M}$ denote strains on the tendon, fibril and
separate molecule \cite{Frat}. 
}
\end{figure}

We shall study the viscoelastic properties of collagen fibrils for a
wide temperature range $20$ to $100 ^{\circ}C$ and at frequencies within
the eigen-frequency domain ($50-20000$ Hz).  The collagen fibril is
composed of collagen triplexes and hydration water with a small amount
of salts.  Fig.~1 shows a simplified scheme of the collagen hierarchy
from the separate triplex to the tendon. Our mechanical methods allow
studying samples with diameter 1 $\mu\,{\rm m}$, and we shall work with
a separate fibril (denoted by F in Fig.~1). See Refs.~\cite{aug} for
optical methods of studying the collagen structure. 

{\it Materials and Methods}.  Achilles tendons of young rats were obtained
from the Yerevan Medical Institute. Separation of fibrils from the fiber
and from each other was carried out mechanically in $96 \%$ of ethyl
alcohol at temperature $5^{\circ}$C using micro-tweezers and microscope.
The experimental sample is a cylinder of length 0.3 mm, which is cut-off
from a separate fibril, held in the micro-tweezers and washed out in
distilled water before experiments. More details on the preparation of
similar experimental samples are found in \cite{Mor1}.

The sample under investigation was enclosed in the experimental chamber
and placed in a temperature-controlled cabinet with the temperature
maintained at $25 ^{\circ}$C. The hydration level of the sample was
adjusted by placing a drop of $CaCl_{2}$ solution at the bottom of the
experimental chamber. The sample was allowed to equilibrate at a given
humidity for several hours. The relative humidities from $97$ to
$32\%$ in the chamber were achieved by means of $CaCl_{2}$ solutions of
different concentrations, while the relative humidities $15\%$ and $10\%$ were obtained
via saturated solutions of $ZnCl_{2}$ and $LiCl$, respectively. The
chamber was then covered by the heat-insulating jacket and placed on the
table of the microscope used to measure the sample vibration. The
viscoelastic properties or the sample length were measured point by
point when varying the temperature continuously at a rate of $1^{\circ}$
C/min. For checking the features of hysteresis and irreversibility (see below)
the heating rate was occasionally reduced to $0.1^{\circ}$C/min.

For measuring the Young's modulus $E$ (defined as the ratio of stress
[pressure] over strain) and the logarithmic damping decrement
$\vartheta$ we applied Morozov's micromethod \cite{Mor2}. The method is
based on the analysis of electrically excited transverse resonance
vibrations of the sample (fibril cylinder), which is cantilevered from
one edge (another edge is free). Modifications of the method that enable
measuring $E$ and $\vartheta$ within a wide temperature range is
described in \cite{Mor1,Mor3}.

\begin{figure}
\includegraphics[width=230pt]{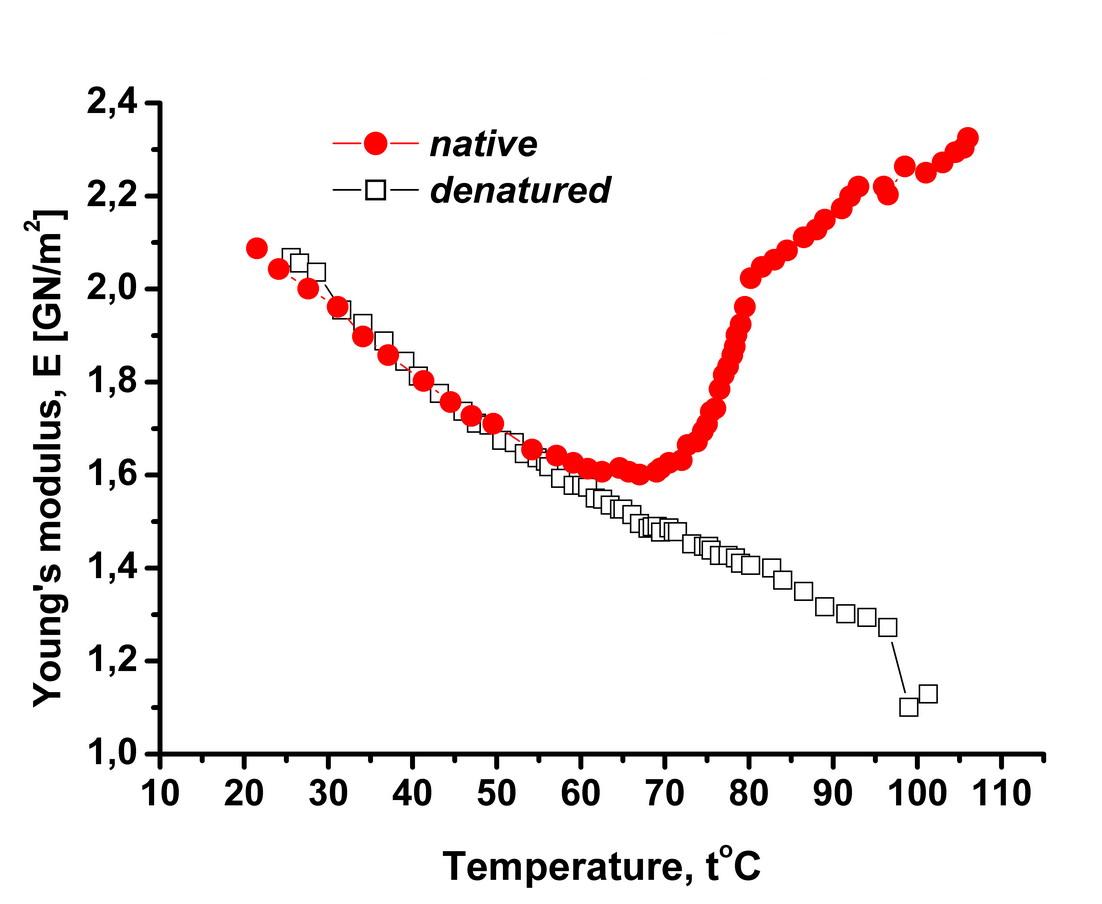}
  \caption{\label{fig2} 
Young's modulus $E$ versus temperature for the native and heat-denatured
collagen fibrils. The heating speed is $1^{\circ} $C/min. Red arrows and indices $1-5$
indicate on temperature intervals discussed in the text.
}
\end{figure}

As a characteristic of internal friction we employ the logarithmic decrement of
damping 
\begin{equation}
\vartheta=\ln[\,{A(t)}/{A(t+T)}\,], 
\end{equation}
where the
oscillation amplitude $A(t)$ of the sample has two consecutive peaks at times $t$ and
$t+T$. For measuring $E$ and the phase-frequency or the amplitude-frequency 
characteristics of oscillations (employed for obtaining $\vartheta$), 
it is necessary to change smoothly the frequency $f$ of the induced oscillations and 
determine the basic resonance frequency, which corresponds to the maximal 
oscillation amplitude of the sample free end. 
Young's modulus for sample main axis is calculated by following formula \cite{Lan}
\begin{equation}
\label{1}
E=3.19\cdot f_{0}^{2}\cdot l^{4}\cdot\rho\cdot P/I_{\rm min}, 
\end{equation}
where $f_0$ is resonance frequency, $l$ is the sample length, $P$ is the
cross-section area, $\rho$ is the density, and $I_{\rm min}$ is the main
inertia moment of that section, which corresponds to the deformation plane
with the minimal stiffness.  For the round cross-section of our samples
$I_{\rm min}={\pi\cdot D^{4}}/{64}$ \cite{Lan} and $P=\pi\cdot D^2/4$,
where $D$ is the sample diameter (measured with precision $0.02\,\mu$m). 
Thus Young's modulus is calculated from
(\ref{1}), where $l$, $\rho$, $P$ and $I_{\rm min}$ are the known sample
characteristics and $f_0$ is measured on the experiment. 

The data on the hydration water content is obtained by measuring the
sample mass via the method of \cite{Gev3}. This method allows to detect
the changes of mass within $0.00001\,{\rm mg}$ in microsamples
weighting up to $0.01\, {\rm mg}$.

\begin{figure}
\includegraphics[width=230pt]{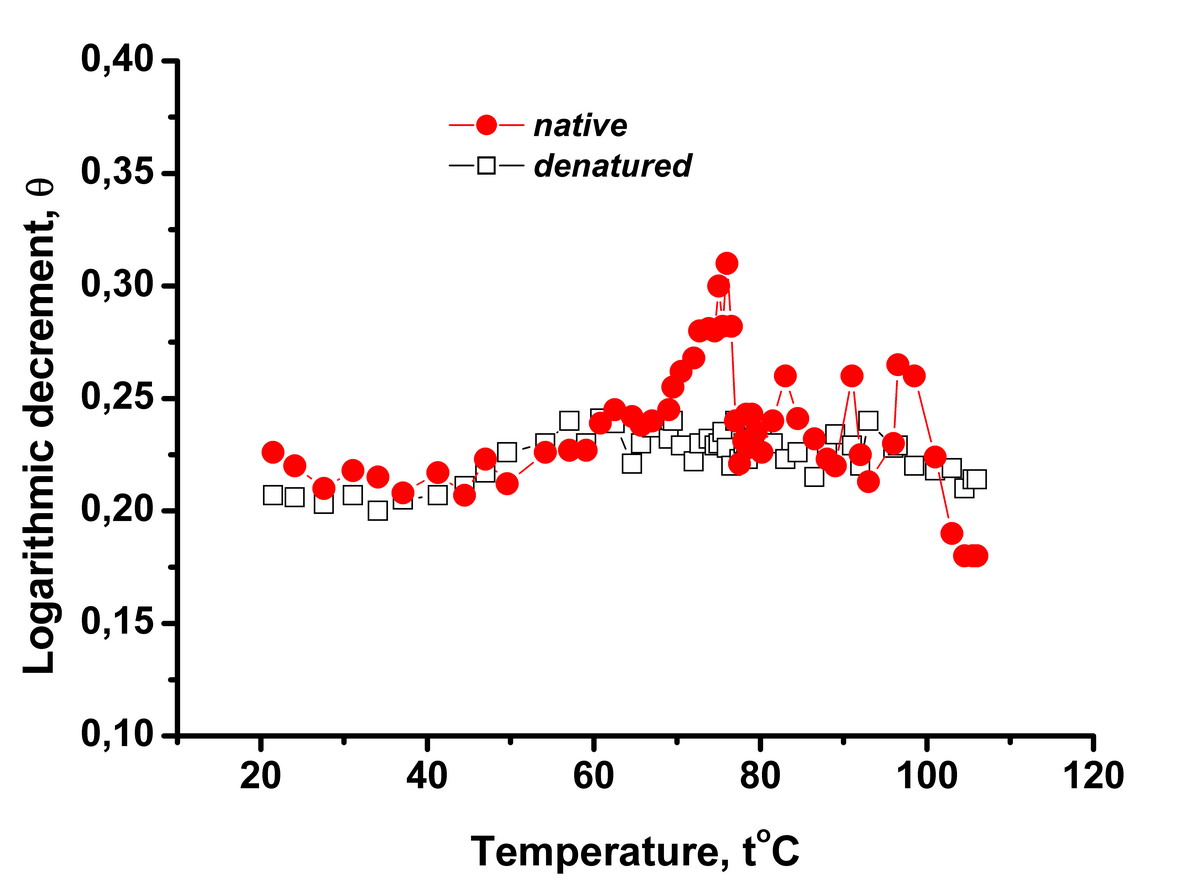}
  \caption{\label{fig3}The logarithmic damping
  decrement $\vartheta$ versus temperature for the native and heat-denatured 
  collagen fibrils.   The heating speed is $1^{\circ}$C/min.
The red arrow and indices $3$ and $4$ correspond to those in Fig.~2.
}
\end{figure}

{\it Results and Discussions}. The main contribution to Young's modulus
of collagen fibril comes from the rigidity of separate collagen
molecules and from the inter-molecular interactions. It was
argued recently that the interplay between these two mechanisms is the
key for understanding the collagen mechanics \cite{buehler_pnas}. 

Fig.~2 displays the Young modulus of the collagen fibril versus
temperature. At the initial temperature $25^\circ$C we created relative
humidity $93\%$. The water content in the fibril is 
$0.3\,{\rm g}\, {\it H_2 O}$/g dry collagen. 

For the considered frequency range ($50-20000$ Hz) the studied
quantities (e.g., Youngs modulus) does not show any significant
dependence on the frequency (not shown on figures).  Thus we are in the
slow deformation regime, e.g., the values of Young's modulus shown in
Fig.~1 are consistent with earler results in this regime
\cite{buehler,gaut_buehler,young}.  It is however expected that
dependence on the frequency will show up for larger frequencies (e.g.,
the Young's modulus starts to increase with frequencies)
\cite{gaut_buehler,buehler}.  For our samples this dependence starts
above $100\,{\rm K}$Hz.  In the studied frequency range the hydrated
water does not contribute directly to the Young's modulus
\cite{Van,Gev3}, but the water can induce structural changes in the
fibril that will alter its elastic features. 

As suggested by our results, the studied temperature domain
$20-100^\circ$C should be separated into five intervals; see Fig.~2.
For each of these intervals we discuss the behavior of the measured
quantities for the native collagen fibril sample and compare it with the
corresponding heat-denaturated sample, which was prepared by keeping the
native sample at $100^\circ$C for $30$ minutes. 

{\bf 1.} Young's modulus of the native collagen smoothly decreases
between $20^\circ$C and $45^\circ$C; see Fig.~2 (temperature borders of
the intervals are defined conventionally). There is no difference between the
Young modulus of the native sample and that of the heat-denaturated sample. 
In this temperature range the
logarithmic decrement of damping (LDD)|which characterizes internal
friction and is generally rather susceptible to inter-molecular
interactions|does not experience any systematic change; see Fig.~3.
The hydrated water content also does not change in this interval; see
Fig.~4. 

Note that the conformational changes in this interval are completely
reversible, since the features of the fibril did not change after
repeating the cooling-reheating process ten times. Though the
Young modulus and the LDD of the native fibril are almost
indistinguishable from those of the heat-denaturated fibril,
the hydrated water content (HWC) does show certain differences between
the native and heat-denaturated sample; see Fig.~3. 

It is likely that in this stage only the triplex conformation changes.
We think that the decrease of Young's modulus in this temperature
interval is a mechanic analogue of very slow single triplex thermal
instability, which was calorimetrically observed in \cite{Leikina}.  In
one way or another, similar instabilities are seen by many experiments
that work at very low concentration, so that the inter-molecular
interactions do not play any role \cite{Priv}. 

{\bf 2.} In the temperature interval $45-58^\circ$C the decrease of
Young's modulus for the native sample is impeded as compared to the
previous stage. Now differences between the native and heat-denaturated
situations set in: the Young modulus of the native sample is larger and
decreases slower as compared to that of the heat-denaturated sample.
Also in this interval we noted first indications of hysteresis and
irreversibility during heating and recooling (not shown on figures). We
checked and confirmed these indications by changing the heating rate
from $1^\circ$C/min to $0.1^\circ$C/min. 

Presumably already in this temperature interval the intermolecular interactions
influence on the change of Young's modulus. Recall that the
collagen triplexes in the fibril are tied with cross-links,
direct hydrogen bonds and water-mediated hydrogen bonds; see Fig.~1~({\bf d}). 

{\bf 3.} The third interval lies in $58-75^\circ$C. Here the Young
modulus of the native sample is nearly constant. To our knowledge such
an effect was never seen for a single collagen triplex.  The Young
modulus of the heat-denaturated sample continues to decreases following
the same linear law as for the previous stages; see Fig.~2.  The
logarithmic decrement of damping (LDD) and the hydrated water content
(HWC) of the native sample increase suddenly. The endpoint of the
interval (approximately $75^\circ$C) plays a special role, since here
Young's modulus starts to increase, while both LDD and HWC assume their
maximal values; see Figs.~3 and 4. In this interval the hysteresis is more
pronounced than for the previous interval. 

Altogether, this seems to indicate that the intermolecular interactions
start to play an important role.  Most likely reason for the sudden
changes of LDD and HWC|which we stress are clearly absent for the
heat-denaturated sample|is that partially molten collagen molecules
start to overlap and create new bonds between each other.  This
facilitates intermolecular interactions.  The hydrated water content
increases, since new adsorption centers open up during the melting. A
similar correlations between the presence of water and the strength of
inter-molecular interactions was obtained in Ref.~\cite{buehler} via
atomistic modeling.

Note that theoretical arguments 
predicted recently that the energy dissipation in the collagen fibril
gets maximized at the transition from homogeneous intermolecular shear
to slip pulses \cite{buehler_pnas}. In our Figs.~2, 3 and 4 we also see
simultaneous indications of structural changes (Young's modulus and
hydrated water content) and dissipation maximization (LDD pick). 

\begin{figure}
\includegraphics[width=230pt]{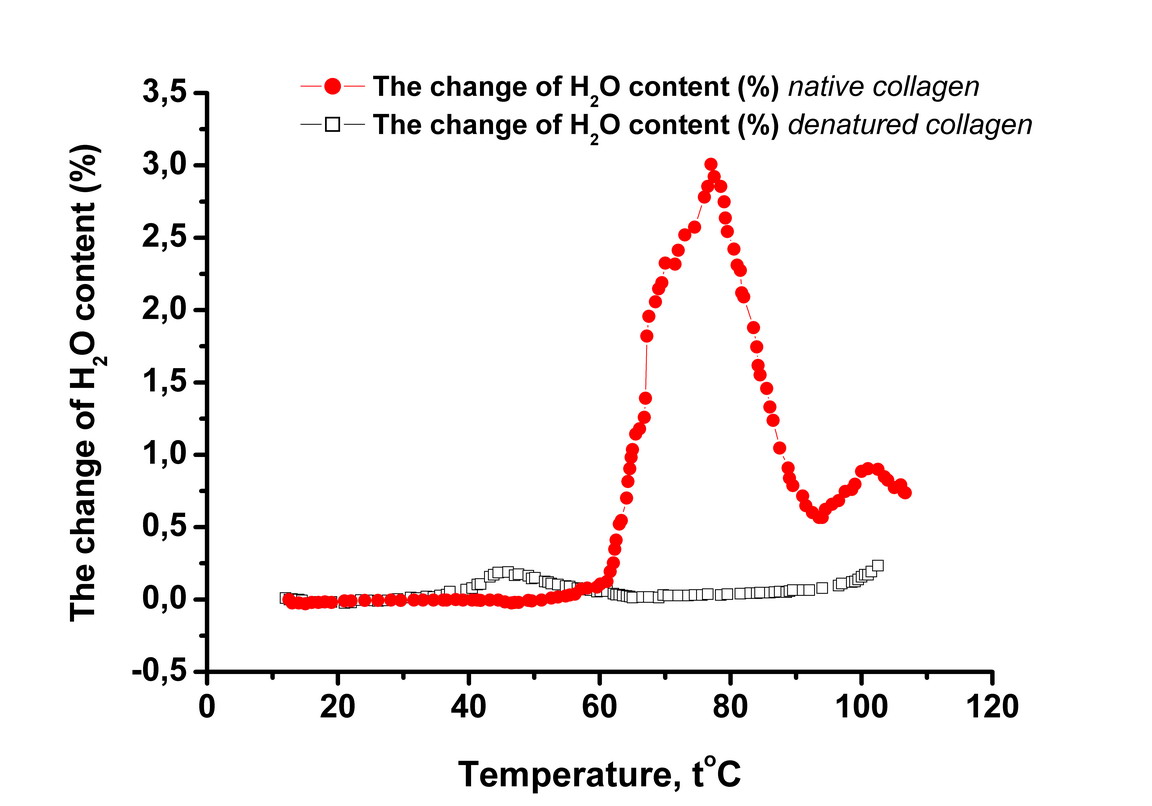}
  \caption{\label{fig4}
The hydrated water content (HWC) versus temperature for
native and heat-denatured collagen fibrils. The initial
temperature and water content in the chamber were $25 ^{\circ}$C and
$93 \%$, respectively. At this temperature and humidity the water content of the
native and denatured micro-fibril is $h = 0.3\, {\rm  g\, {\it H_{2}O} / g\,
dry\, collagen}$ and $h = 0.22 {\rm \,g\, {\it H_{2}O} / g\, dry\, collagen}$,
respectively. The change of the water content is given in percents
relative to that water content at $25 ^{\circ}$C. The heating speed
is $1^{\circ}$ C/min. }
\end{figure}

{\bf 4.} Between $75-80^\circ$C Young's modulus of the native sample
starts to increase in sharp contrast to Young's modulus of the
heat-denaturated sample that keeps decreasing; see Fig.~1.
Simultaneously, both LDD and HWC start to decrease; see Figs.~3 and 4.
At the end of this interval (i.e., at $80^\circ$C) Young's modulus
almost approaches its initial value at $25^\circ$C. 

We think that a possible reason for increasing Young's modulus is that
the network of the inter-molecular bonds (established already during the
previous stage) develops and contributed significantly to the rigidity.
This can also expel the water out of the fibril. 

We see that the crucial difference between the measured characteristics
of the native and heat-denaturated sample indicate on the existence of
an important structural feature of the native fibril, which ensures its
stability for temperatures larger than $58^\circ$C and which is absent
for the heat-denaturated sample. 

{\bf 5.} Above $80^\circ$C the Young modulus keeps on increasing, though
slower than for the previous step. The LDD stops decreasing and starts
to change irregularly, in contrast to the LDD of the heat-denaturated
sample. The HWC in this region keeps on decreasing before $92^\circ$C,
and then changes non-monotonously.  The dynamics in this region is
irreversible: if the heating stops at some temperature larger
than $80^{\circ}$C and the sample is cooled back to $20^\circ$C, the
above behavior of the native fibril is not recovered upon subsequent
heating. Instead we obtain the heat-denaturated behavior displayed on
Figs.~2--4. On the other hand, if the heating simply stops at some
temperature larger than $80^{\circ}$C, the sample slowly relaxes to the
heat-denaturated value of the Young modulus. 

It is likely that the origin of this irreversibility is mainly entropic:
there are already so many well-established inter-molecular bonds in this
regime that the reverse transition to the weakly coupled inter-molecular
situation becomes impossible. The HWC and LDD in this interval behave
non-monotonously indicating on further structural changes, which are
again absent for the heat-denaturated sample. 

{\it In conclusion}, we studied thermal stability of the (type I)
collagen fibril via measuring its Young's modulus, logarithmic decrement
of damping (LDD) and hydrated water content (HWC).  All the measurements
were done in parallel for the native collagen fibril from rat's tendon
and its heat-denaturated version. We aimed to understand how the
instability of a single collagen triplex is assimilated by this
structure.  We observed that between $20^\circ$C and $50^\circ$C the
Young modulus of the native fibril decays with increasing the
temperature.  This indicates on the (partial) instability of the fibril
due to single-collagen effects. For higher temperatures the LDD and HWC
indicate on serious structural changes in the fibril. Due to these
changes Young's modulus first becomes constant and then (upon further
heating) increases with temperature. None of these effects is
seen on the heat-denaturated fibril, which displays monotonously
decaying Young's modulus and a relatively trivial behavior for LDD and
HWC. 

This work was supported by Auburn University Detection and Food Safety
Center, NSF (Grant CTS-0330189 to ALS) and Volkswagenstiftung (to AEA).


\end{document}